\documentclass[aps,prl,twocolumn]{revtex4}
\usepackage{graphicx,epsfig,amsmath,amssymb,amsfonts,mathrsfs,bm,txfonts}
\begin{document} 
\title{Nonequilibrium electron spectroscopy of Luttinger liquids}
\author{So Takei, Mirco Milletar\`i, and Bernd Rosenow}
\affiliation{Max-Planck-Institut f\"ur Festk\"orperforschung, D-70569 Stuttgart, Germany}
\date{\today}

\begin{abstract}
Understanding the effects of nonequilibrium on  strongly  interacting quantum systems is a challenging 
problem in condensed matter physics. In dimensions greater than one,  interacting electrons can often be 
understood within Fermi-liquid theory where low-energy excitations are weakly interacting quasiparticles. 
On the contrary, electrons in one dimension are known to form a strongly-correlated phase of matter called 
a Luttinger liquid (LL), whose low-energy excitations are collective density waves, or plasmons, of the electron 
gas. Here we show that spectroscopy of locally injected high-energy electrons can be used to probe energy 
relaxation in the presence of such strong correlations. For detection energies near the injection energy, 
the electron distribution is described by a power law whose exponent depends in a continuous way on
the Luttinger parameter, and energy relaxation can be attributed to plasmon emission. For a chiral  LL 
as realized at the edge of a fractional quantum Hall state, the distribution function grows linearly with the 
distance to the injection energy, independent of filling fraction. 
\end{abstract}
\maketitle

Over the last decade, experimental advances in nanostructure fabrication have brought a resurgence of 
interest in the LL model because of the possibility to test its peculiar predictions 
\cite{ mandl,giamarchi,Wen1990,Wen1990-2,KaFi92}. Defining signatures of a LL such as spin-charge separation 
\cite{orgcond,scs}, charge fractionalization \cite{chargefrac,chargefrace}, and the power-law suppression of the 
local electron tunneling density of states \cite{swcne1,swcne2,fqhee1,fqhee2,fqhee3} have been experimentally 
verified. Recently, LLs driven far from equilibrium have begun to receive attention  
\cite{tsexp,inhomoint2,disorder1,khodas07}. Studying these systems offers the possibility to characterize novel aspects 
of electron-electron interactions and to understand energy relaxation processes that have not been apparent in 
the above-mentioned equilibrium experiments.

Here we consider a LL driven out of equilibrium by local injection of high-energy electrons, far away from any contacts, 
at a fixed energy. Their spectral properties are extracted at another spatial point some distance away by evaluating the 
average tunneling current from the LL into an empty resonant level with tunable energy. In this work, we consider both  
standard (non-chiral) and chiral LLs, which are realized at the edge of fractional quantum Hall systems 
\cite{Wen1990,Wen1990-2,fqhee1,fqhee2,fqhee3}. 

For the standard LL and for probe energies slightly below the injection energy, we find that the inelastic component 
of the current shows a power law behavior as a function of the  difference between injection and detection energy, 
with an exponent that continuously evolves as the interaction parameter is varied. We develop a perturbative 
approach which shows how injected electrons can relax by emitting plasmons inside the wire.

For a chiral LL at  the  edge of a fractional quantum Hall state from the Laughlin sequence, an essentially exact 
calculation of the tunneling current is possible. Here, the inelastic part of the electron current {\em increases} in a 
linear fashion as the probe energy is lowered from the injection energy towards the chemical potential of the 
edge state, despite a {\em decreasing} tunneling density of states for electrons. This behavior is compatible 
with our result for the standard LL in the limit of strongly repulsive interactions. For probe energies close to the 
chemical potential, the chiral LL is far from equilibrium: the electron spectral function approaches a finite value, 
in striking contrast to the power law decrease towards zero in equilibrium. In addition to the inelastic contribution 
to the probe current, in a chiral LL there always is an elastic contribution, indicating that a finite fraction of electrons 
travels from the injection to the probe site without loosing energy. 

\begin{figure}[t]
\begin{center}
\includegraphics[scale=0.5]{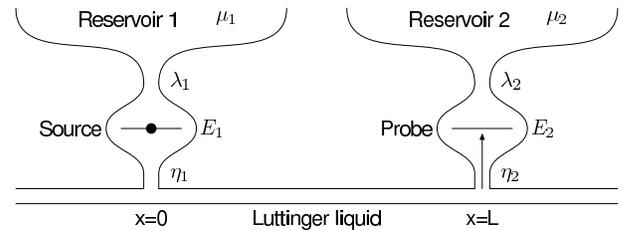}
\caption{\textbf{The proposed experimental setup}. Hot electrons are injected from the source resonant level
at $x=0$, and are collected at the probe resonant level at $x=L$. System parameters are set (see text)
so that the source (probe) occupancy is fixed to be full (empty). Spectral properties of the
injected electrons are extracted by measuring the tunneling current between the edge and the probe
(indicated by the arrow).}
\label{setup}
\end{center}
\end{figure}

Electrons with charge $e_0$ are injected into the  LL from a resonant level (source) with energy 
$E_1\equiv e_0 V_1>0$ at position $x=0$ (see Fig.\ref{setup}). Energy relaxation is studied by coupling a 
second resonant level (probe) with energy $E_2\equiv e_0 V_2>0$ to the LL at position $x=L$ (downstream  
for the chiral LL), and by computing the tunneling electron current between the LL and that level. The two levels 
are coupled to the LL via tunneling amplitudes $\eta_1$ and $\eta_2$, respectively. In addition, source 
and probe dots are  coupled to reservoirs held at chemical potentials $\mu_1$ and  $\mu_2$ via tunneling 
amplitudes $\lambda_1$ and  $\lambda_2$.  The chemical potential of the LL is taken to be zero. We assume 
the level broadening due to tunnel couplings to be small in comparison to both $E_1$ and $E_2$, and  
therefore consider the current in the sequential-tunneling regime. Further,  we assume 
$\lambda_1\gg\eta_1$ with $\mu_1>E_1$ so that the source occupancy is constrained 
to one, and   $\lambda_2\gg\eta_2$ with $\mu_2<E_2$ so that the probe occupancy is fixed
at zero.

\begin{figure}[t]
\begin{center}
\includegraphics[scale=0.34]{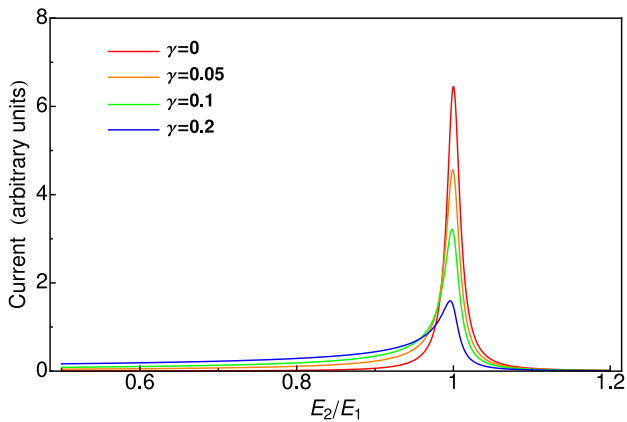}
\caption{\textbf{Tunneling current for the standard case for various interaction}. 
The current is plotted for zero temperature and includes the leading contribution in $\Delta E/E_1$. 
The inelastic contribution for $E_2<E_1$ shows a power law decay as a function of increasing $\Delta E$
with an exponent that depends on the interaction parameter (see text). A level broadening of $0.01E_1$
is used for the elastic peak.}
\label{res2}
\end{center}
\end{figure}
We first focus on the standard LL and consider spinless electrons, for which the interaction strength is 
described by a single parameter $K$ \cite{giamarchi}. The case $K=1$ describes non-interacting electrons, 
$K<1$ corresponds to repulsive interactions, and $K>1$ to attractive ones. We use the non-equilibrium 
Keldysh formalism \cite{RandS} to calculate the current flowing into the probe dot to leading order in the tunneling 
amplitudes $\eta_1$ and $\eta_2$, details are described in the appendix. 
\begin{equation}
\label{Inonchiral}
I=-\frac{2\pi e_0}{\hbar}\frac{|\eta_1|^2|\eta_2|^2\theta(\Delta E)}{u^2\hbar^2E_1\Gamma^2(1+\gamma)}
\left(\frac{\alpha E_1}{u\hbar}\right)^{4\gamma}
\left[\frac{(\Delta E/E_1)^{2\gamma-1}}{\Gamma(2\gamma)}\right]\ .
\end{equation}
Here,  $\Gamma(x)$ is the gamma-function, $u$ is the velocity of plasmon excitations, $\alpha$ denotes the short 
distance cutoff of the theory, and $\gamma=K(1/K-1)^2/4\ge 0$. When calculating the current Eq.~(\ref{Inonchiral}), the 
limit of large interdot separation $u/L \ll E_1, E_2$ was taken. In the non-interacting limit ($\gamma\rightarrow 0$) the 
quantity in the square brackets is a representation of the delta-function, and Eq.~(\ref{Inonchiral}) reduces to 
$I\propto\delta(\Delta E)$.  When the interactions are turned on, the elastic peak gradually broadens to give rise to an inelastic 
contribution which shows a power law decay as a function of increasing $\Delta E$, with an 
exponent that continuously evolves as a function of the interaction parameter. For strong enough interactions with 
$\gamma > 1/2$, the elastic peak vanishes and the remaining inelastic contribution monotonically increases with a power 
law which again evolves as a function of the interaction parameter. The result Eq.~(\ref{Inonchiral}) is plotted in 
Fig.\ref{res2}. Broadening of the peak is included in the figure to reflect the finite width of the resonant levels due to the 
couplings to the reservoirs and the wire.

\begin{figure}[t]
\begin{center}
\includegraphics[scale=0.365]{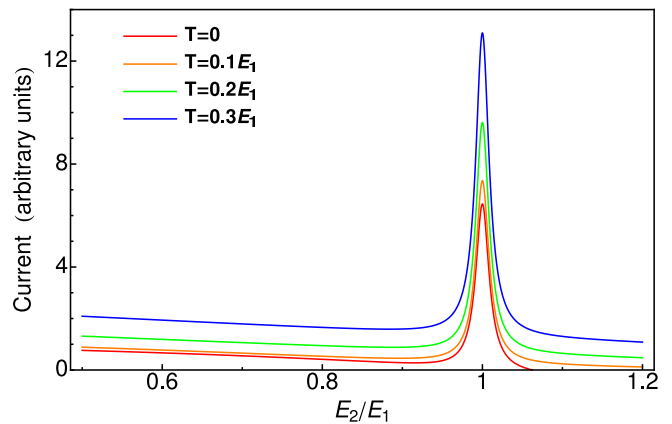}
\caption{\textbf{Tunneling current for the chiral case at various temperatures}. The current shows an elastic 
contribution at $E_1=E_2$, and an inelastic contribution for $E_2<E_1$ which increases as energy transfer is 
increased. The broadening of the elastic peak is included as in the standard case. The same level broadening 
as the standard case is used here.}
\label{res}
\end{center}
\end{figure}

In the limit of weak interactions with Luttinger parameter $K$ close to one, energy relaxation as described by  
Eq.~(\ref{Inonchiral}) can be interpreted by using lowest order perturbation theory in the interaction strength. 
Interactions can be decomposed into forward scattering between electrons near the same Fermi point with amplitude 
$g_4$, and  between electrons near opposite Fermi points with amplitude $g_2$. The $g_4$-interaction merely renormalizes 
the fermion and plasmon velocities and cannot give rise to relaxation. In a spatially homogeneous LL, 
the $g_2$-process cannot give rise to energy relaxation either due to  the simultaneous requirement 
of momentum and energy conservation. However, because of the local nature of  injection and collection 
processes considered here, an electron is capable of exploring virtual momentum states in connection with 
tunneling, and a consecutive inelastic process can both conserve momentum and produce a final state 
with the same total energy as the source state. Here, we consider the lowest order inelastic process 
proportional to $\gamma\propto g_2^2$ at zero temperature, in which an electron is transported from 
the source to the probe while emitting a single plasmon inside the wire. First, the electron in the source 
tunnels into a right moving momentum eigenstate whose energy may be different from $E_1$.  In the 
second step, a left moving plasmon with energy $\Delta E$ is emitted via a $g_2$-process, and
the right moving electron is scattered into another wire state such that momentum is  conserved.
After propagation along the wire,  the electron tunnels into the probe. Alternatively, tunneling into the wire can be elastic, 
and the plasmon can be emitted when tunneling into the probe. One finds that the matrix element for these processes 
scales as $1/\sqrt{|\Delta E|}$. This can be understood by multiplying the matrix elementfor plasmon emission, which  
increases as $\sqrt{|\Delta E}|$, with the time available for plasmon  emission, which diminishes as $1/|\Delta E|$ due 
to the energy-time uncertainty principle. The tunnel current can then be computed using Fermi's golden rule, and correctly 
reproduces the inelastic component $I \propto \gamma/\Delta E$ of equation (\ref{Inonchiral}) to order $\gamma$.

Next, we consider tunneling into a chiral LL at the edge of a fractional quantum Hall state from the Laughlin sequence.
We focus on the filling fraction  $\nu = 1/3$, where the area occupied by one electron is threaded by three quanta of magnetic 
flux. The calculation of the steady state current proceeds along the same lines as for Eq.~(\ref{Inonchiral}), with the 
difference that we were able to obtain an exact expression for all values of $\Delta E$ and for finite temperature, 
\begin{widetext}
\begin{equation}
I =  -e_0 \frac{\pi^3|\eta_1|^2|\eta_2|^2\alpha^4(k_BT)^3}{4u^6\hbar^7}  \left[ \frac{X_1^2 e^{\frac{X_1}{2}}}
{\cosh{(X_1/2)}} \left(1+\frac{X_1^2}{\pi^2}\right) \delta(\Delta X)+
\frac{3\Delta Xe^{\Delta X/2}}{\sinh(\Delta X/2)} 
\sum_{i=1}^2\frac{e^{X_i/2}}{\cosh{(X_i/2)}}\left(1+\frac{X_i^2}{\pi^2}\right)\right]. 
\label{finiteT.eq}
\end{equation}
\end{widetext}
Here,  $\Delta X = X_1-X_2$, $X_i= E_i/k_B T$. At zero temperature, the expression for the current simplifies 
to $I \propto   E_1^4 \delta(\Delta E) +6 \theta(\Delta E) (E_1^2+E_2^2)\Delta E$ where $\Delta E=E_1-E_2$.
The current, plotted for both zero and finite temperatures in Fig.\ref{res}, has two main contributions:
elastic and inelastic. The peak is due to electrons that were elastically transported from the source to the 
probe. Second, there is a broad inelastic contribution that extends over the range $E_2 < E_1$, and that 
grows monotonically as $E_2$ is lowered. For $E_2 \lesssim E_1$, the current increases linearly with 
$\Delta E$. We have confirmed that a similar inelastic contribution to the current is also present for the 
Laughlin filling fraction  $\nu=1/5$. In this case, an exact computation at zero temperature shows again that 
$I_{\rm inel}\propto\Delta E$ for $E_2\lesssim E_1$. This  suggests that the linear upturn in the current below 
$E_1$ may be a generic feature at all Laughlin filling fractions.

For a non-interacting chiral Fermi liquid, which describes the edge excitations of an integer quantum Hall state, 
hot electrons do not relax. In addition, the weight of the elastic peak is reduced as the temperature is increased. 
This reduction is due to Pauli blocking of states by thermally excited edge electrons residing above the chemical
potential. When interactions are present, Fig.\ref{res} shows an overall increase in the elastic 
current with temperature. This reflects the increase in the tunneling density of states with temperature 
and constitutes a clear signature of LL physics. 

The setup of Fig.\ref{setup} is ideal for directly extracting the electron energy distribution, $f(E)$,
and spectral function, $A(E)$, inside the wire at a spatial point far from the injection site. With the probe occupancy 
constrained to be empty,  the tunneling current is given by $I_{\rm empty}=ie_0|\eta_2|^2G^<(E)$, while a similar 
evaluation with probe occupation held full gives $I_{\rm full}=ie_0|\eta_2|^2G^>(E)$ \cite{ChaWen95}. Once the two 
currents are obtained, both $f(E)$ and $A(E)$ can be extracted by expressing  the lesser and greater Green functions, 
$G^<(E)=if(E)A(E)$ and $G^>(E)=-i(1-f(E))A(E)$, in terms of electron distribution function and spectral weight. 
At zero temperature and for $\nu=1/3$, $f(E_2)$ and $A(E_2)$ valid for $0<E_2<E_1$ read
\begin{equation}
\label{ae}
A(E_2)=\frac{\alpha^2}{2u^3\hbar^4}\left[E_2^2 +\left(E^*\right)^2
\frac{\Delta E}{E_1}\right],
\end{equation}
\begin{equation}
\label{fe}
f(E_2)=\frac{\left[1+\left(\frac{E_2}{E_1}\right)^2\right]\frac{\Delta E}{E_1}}
{\left(\frac{E_2}{E^*}\right)^2+\frac{\Delta E}{E_1}}, 
\end{equation}
where $E^*=\sqrt{6\pi|\eta_1|^2\alpha^2E_1^3/u^3\hbar^3}$ separates two energy regimes. We note 
that $E^*$ can be parametrically larger than the level widths such that our sequential tunneling approximation stays 
valid. In the high-energy regime and for small energy transfers 
($E^*\ll E_2\lesssim E_1$), $f(E_2)\approx 12\pi|\eta_1|^2\alpha^2\Delta E/u^3\hbar^3$, which shows 
that the linear upturn in the current below $E_1$ is also reflected in the distribution function. In the same regime, 
we find that the spectral function does not deviate strongly from its equilibrium expression (with $\eta_1=0$). In the 
low-energy regime ($0\lesssim E_2<E^*$), $f(E_2)$ smoothly approaches one and the spectral function approaches a 
finite value. The latter is in stark contrast to the equilibrium case. 

\textbf{Acknowledgment}: 
We  thank A.~Yacoby for drawing our attention to the problem of local and energy 
resolved  electron injection, and W.~Metzner and V.~Venkatachalam for useful discussions.  
 B.~R.~was supported by  the Heisenberg program of DFG.

\textbf{Appendix}: 
We now provide the theoretical basis for the derivation of Eqs.~(\ref{Inonchiral},\ref{finiteT.eq}). 
The system is modeled by the Hamiltonian $H= H_{\rm LL} + H_{\rm dot}+H_{\rm tun}$, where 
$H_{\rm LL}$ models the LL, $H_{\rm dot} = E_1 \psi_1^\dagger \psi_1 
+ E_2 \psi_2^\dagger \psi_2$ the two resonant states, and $H_{\rm tun}$ describes the tunneling of 
electrons between the wire and the two resonant levels. $\psi_1$ ($\psi_2$) are electron 
operators of the source (probe) with occupation numbers $\langle \psi_1^\dagger \psi_1\rangle =1$ 
and $\langle \psi_2^\dagger \psi_2\rangle =0$. The standard LL Hamiltonian reads \cite{giamarchi} 
\begin{equation}
H_{\rm LL}=\frac{u}{4\pi K}\int dx  [(\partial_x\phi_R(x))^2+(\partial_x\phi_L(x))^2].
\label{hamiltonian.eq}
\end{equation}
where $K$ is the LL parameter,  and the left and right moving boson operators satisfy $[\phi_R(x),\phi_R(x')]
=-[\phi_L(x),\phi_L(x')]=i\pi K\,\mbox{sgn}(x-x')$. One-dimensional electron densities are given by
$\rho_{R,L}(x)=(\partial_x\phi_{R,L}(x))/2\pi$ and $u$ denotes the plasmon velocity. To simplify the notation, 
we use the units where $\hbar=1$ and $k_B=1$. The tunneling Hamiltonian is given by
\begin{equation} \label{tunn}
H_{\rm tun} = \eta_1\psi_{1} \psi^\dag(x=0)  + \eta_2\psi_2 \psi^\dagger(x=L)  +h.c. \ \ .
\end{equation}
where $\psi(x)=\psi_R(x)+\psi_L(x)$. The electron operators can be bosonized as 
$\psi_{R,L}(x)=\mbox{exp}[i(K_\pm\phi_R(x)+K_\mp\phi_L(x))]/\sqrt{2\pi\alpha}$ with $K_\pm=(K^{-1}\pm 1)/2$.
The expectation value of the current reads
\begin{equation} \label{exp}
I = \langle T_c \lbrace \hat{I}_{\rm cl}(t_1)e^{-i\int_c d t H_{\rm tun}(t)} \rbrace \rangle_0,
\end{equation}
where all operators are written in the interaction picture with respect to $H_{\rm LL} + H_{\rm dot}$.
The current is computed using the nonequilibrium Keldysh technique \cite{RandS}, and 
$T_c$ indicates time-ordering of the operators on the time-loop contour $c$. The ``classical" 
component of the current operator is the symmetric combination of the operator on the 
forward ($+$) and backward ($-$) parts of the Keldysh contour, i.e. $\hat{I}_{\rm cl}(t)=(\hat{I}_{+}(t)+\hat{I}_{-}(t))/2$, where 
$\hat{I}_{\pm}(t)=-ie_0[H_\pm,\psi^\dag_{2,\pm}\psi_{2,\pm}]$. Upon imposing the constraints on the resonant level occupancies 
and taking the limit of large inter-dot separation, the time of propagation $L/u$ drops out and we arrive at the following 
expression for the steady state current to leading order in $\eta_1$ and $\eta_2$,
\begin{widetext}
\begin{equation} \label{currLL}
I = e_0 |\eta_1|^2|\eta_2|^2 \int d^3t e^{-i E_2 t_2
+i E_1 t_{34} }(iG^<_{2\gamma+1}(-t_{2}))(iG^>_{2\gamma+1}(t_{34}))
\lbrace \Pi^{<>}_{1+\gamma}-\Pi^{<<}_{1+\gamma}+\Pi^{<>}_{\gamma}
-\Pi^{<<}_{\gamma}+2\Pi^{<>}_{\gamma'}-2\Pi^{<<}_{\gamma'} \rbrace.
\end{equation}
\end{widetext}
Here,  $\gamma=K_-^2K$, $\gamma'=(K_-^2+K_-)K$, and $t_{ij}=t_i-t_j$. The ordering on the Keldysh contour for a 
related problem is described in \cite{K&F}. The factors of correlation functions,
\begin{equation} \label{g}
iG^{\stackrel{>}{<}}_{\beta}(t)= \pm\frac{1}{2 \pi \alpha} \frac{(\pi T \alpha /u)^{\beta}}
{[\sin \pi T (\alpha/u \pm i t)]^{\beta}},
\end{equation}
can be interpreted as the tunneling in and out density of states, and the $\Pi$-matrices, 
\begin{equation} \label{k}
\Pi^{\rho \sigma}_{\beta}=\frac{G^{\rho}_{\beta}(t_{23})G^{\sigma}_{\beta}(-t_{4})}{ G^{\rho}_{\beta}(-t_{3})G^{\sigma}_{\beta}(t_{24})},
\end{equation}
describe the propagation of electrons along the wire. In principle, the current in equation (\ref{exp}) contains 
another term proportional to only $|\eta_2|^2$ that describes the tunneling of electrons into the probe from thermal 
excitations in the wire. However, for low temperatures ($T\ll E_1,E_2$), this contribution is exponentially suppressed. 

The current for the chiral LL Eq.~(\ref{finiteT.eq}) can be derived in a similar fashion. 
First, we note that the Hamiltonian is analogous to that in Eq.~(\ref{hamiltonian.eq}), but with only 
one boson field, say  $\phi_R$, and with $K$ replaced by the filling fraction $\nu$. 
The tunneling Hamiltonian is identical to equation (\ref{tunn}), and the electron operator is
now bosonized as  $\psi(x)=e^{i \phi_R(x)/\nu}/\sqrt{2\pi\alpha}$. The formal expression for the current is still given 
by equation (\ref{exp}). Using similar steps as above, one finds
\begin{multline} \label{curr2}
I = e_0 |\eta_1|^2|\eta_2|^2 \int_{-\infty}^{\infty} dt_2 dt_3 dt_4 e^{-i E_2 t_2
+i E_1 t_{34} }\\
\times(iG^<_{1/\nu}(-t_{2}))(iG^>_{1/\nu}(t_{34})) \lbrace \Pi^{<>}_{1/\nu}-\Pi^{<<}_{1/\nu}
\rbrace.
\end{multline}
The correlation functions and the $\Pi$-matrices are again given by equations (\ref{g},\ref{k}).


\end{document}